\documentclass[
 reprint,
superscriptaddress,
 amsmath,amssymb,
 aps,
superscriptaddress,
 amsmath,amssymb,
 aps,longbibliography
]{revtex4-1}

\usepackage{graphicx}
\usepackage{dcolumn}
\usepackage{bm}
\usepackage{amsmath,amssymb,amstext,amsthm}
\usepackage{braket}
\usepackage{bbold}
\usepackage{bbm}
\usepackage{xcolor}
\usepackage{ gensymb }
\usepackage{soul}
\setstcolor{red}
\usepackage{ mathrsfs }
\usepackage{cancel}

\newcommand{\sigperpSqr}[0]{\sigma^2_{_{\mkern-6mu\perp}}}

\newcommand{\sigperp}[0]{$\sigma_{_{\mkern-6mu\perp}}$}

\newcommand{\sigperpEQ}[0]{\sigma_{_{\mkern-6mu\perp}}}

\begin{document}

\preprint{APS/123-QED}
\setlength{\abovedisplayskip}{1pt}
\title{Small-angle scattering interferometry with neutron orbital angular momentum states}

\author{Dusan Sarenac}
\email{dusansar@buffalo.edu}

\affiliation{Department of Physics, University at Buffalo, State University of New York, Buffalo, New York 14260, USA}
\affiliation{Institute for Quantum Computing, University of Waterloo,  Waterloo, ON, Canada, N2L3G1}

\author{Melissa E. Henderson} 
\affiliation{Institute for Quantum Computing, University of Waterloo,  Waterloo, ON, Canada, N2L3G1}
\affiliation{Department of Physics and Astronomy, University of Waterloo, Waterloo, ON, Canada, N2L3G1}
\author{Huseyin Ekinci} 
\affiliation{Institute for Quantum Computing, University of Waterloo,  Waterloo, ON, Canada, N2L3G1}
\affiliation{Department of Physics and Astronomy, University of Waterloo, Waterloo, ON, Canada, N2L3G1}
\author{Charles W. Clark}
\affiliation{Joint Quantum Institute, National Institute of Standards and Technology and University of Maryland, College Park, Maryland 20742, USA}
\author{David G. Cory}
\affiliation{Institute for Quantum Computing, University of Waterloo,  Waterloo, ON, Canada, N2L3G1}
\affiliation{Department of Chemistry, University of Waterloo, Waterloo, ON, Canada, N2L3G1}

\author{Lisa DeBeer-Schmitt} 
\affiliation{Neutron Scattering Division, Oak Ridge National Laboratory, Oak Ridge, TN 37831, USA}

\author{Michael G. Huber}
\affiliation{National Institute of Standards and Technology, Gaithersburg, Maryland 20899, USA}

\author{Owen Lailey} 
\affiliation{Institute for Quantum Computing, University of Waterloo,  Waterloo, ON, Canada, N2L3G1}
\affiliation{Department of Physics and Astronomy, University of Waterloo, Waterloo, ON, Canada, N2L3G1}

\author{Jonathan S. White}
\affiliation{Laboratory for Neutron Scattering and Imaging, Paul Scherrer Institut, CH-5232 Villigen PSI, Switzerland}

\author{Kirill Zhernenkov}
\affiliation{J\"ulich Centre for Neutron Science at Heinz Maier-Leibnitz Zentrum, Forschungszentrum J\"ulich GmbH, 85748 Garching, Germany}

\author{Dmitry A. Pushin}
\email{dmitry.pushin@uwaterloo.ca}
\affiliation{Institute for Quantum Computing, University of Waterloo,  Waterloo, ON, Canada, N2L3G1}
\affiliation{Department of Physics and Astronomy, University of Waterloo, Waterloo, ON, Canada, N2L3G1}

\date{\today}


\pacs{Valid PACS appear here}


\begin{abstract}

Access to the neutron orbital degree of freedom has been enabled by the recent actualization of methods to prepare and characterize neutron helical waves carrying orbital angular momentum (OAM) at small-angle neutron scattering (SANS) facilities. This provides new avenues of exploration in fundamental science experiments as well as in material characterization applications. However,  it remains a challenge to recover phase profiles from SANS measurements. We introduce and demonstrate a novel neutron interferometry technique for extracting phase information that is typically lost in SANS measurements. An array of reference beams, with complementary structured phase profiles, are put into a coherent superposition with the array of object beams, thereby manifesting the phase information in the far-field intensity profile. We demonstrate this by resolving petal-structure signatures of helical wave interference for the first time: an implementation of the long-sought recovery of phase information from small-angle scattering measurements. 
\end{abstract}
\maketitle

\section{Introduction}

Small-angle neutron scattering (SANS) is a versatile technique that is used to probe the nanoscale structure and dynamics of materials such as polymers, biomolecules, magnetic nanoparticles, porous media, and metal--organic frameworks~\cite{schmitt2020mesoporous,perera2018small,lai2019silicon,heller2023small,draper2020using,heller2021small,Allen:uu5002}. Applications of SANS include studying self-assembly, phase diagrams, interactions of soft matter systems, investigating the magnetic properties and functionalization of nanomaterials, and combining SANS with neutron imaging to access multiple length scales from micrometers to nanometers~\cite{fuhrman2015interaction,qian2018new,kezsmarki2015neel,nagaosa2013topological,muhlbauer2019magnetic}. SANS is complementary to other scattering methods such as small-angle X-ray scattering (SAXS), and can benefit from the use of contrast variation and isotope labelling~\cite{schmatz1974neutron,jeffries2021small,hollamby2013practical}.

In a typical SANS measurement neutrons propagate for several meters after passing through the sample and its environment before they are detected by a position--sensitive neutron detector. Phase information is sacrificed in detection since the far-field neutron intensity profile is proportional to the square modulus of the Fourier transform of the outgoing wave function. Recent work, inspired by computed tomography methods, examined phase-recovery by employing multi-angle SANS measurements~\cite{heacock2020neutron,henderson2023three}.

The need for phase retrieval in SANS is heightened by the developments in production and detection of structured neutron waves, and their use as probes of material properties~\cite{clark2015controlling,sarenac2016holography,sarenac2019generation,larocque2018twisting,geerits2023phase,sarenac2018methods,nsofini2016spin,jach2022method,sherwin2022scattering,afanasev2021elastic,afanasev2019schwinger,sarenac2022experimental,le2023spin}. These techniques rely on tailoring the transverse phase profile of the wavefront to induce non-trivial propagation characteristics such as orbital angular momentum (OAM), non-diffraction, and self-acceleration.~\cite{rubinsztein2016roadmap,bliokh2023roadmap,ivanov2022promises,chen2021engineering,ni2021multidimensional}. For example, imprinting an azimuthal phase gradient upon the wave function prepares helical waves that carry OAM and display a doughnut--like intensity profile in the far-field.~\cite{LesAllen1992,Bazhenov1990,berry2004optical}.

To advance SANS techniques for disambiguating the phase profile of the wave function, we adopt concepts from neutron interferometry~\cite{Sam}. We create a coherent superposition of structured object and reference beams, thereby yielding observable interference amplitudes to reveal phase information in the far field. We demonstrate this method by observing the petal-structure interference between two helical beams that have been prepared using arrays of fork-dislocation phase-gratings, and whose transverse amplitude profile takes on a doughnut-like form with propagation. 

\begin{figure*}
    \centering\includegraphics[width=.85\linewidth]{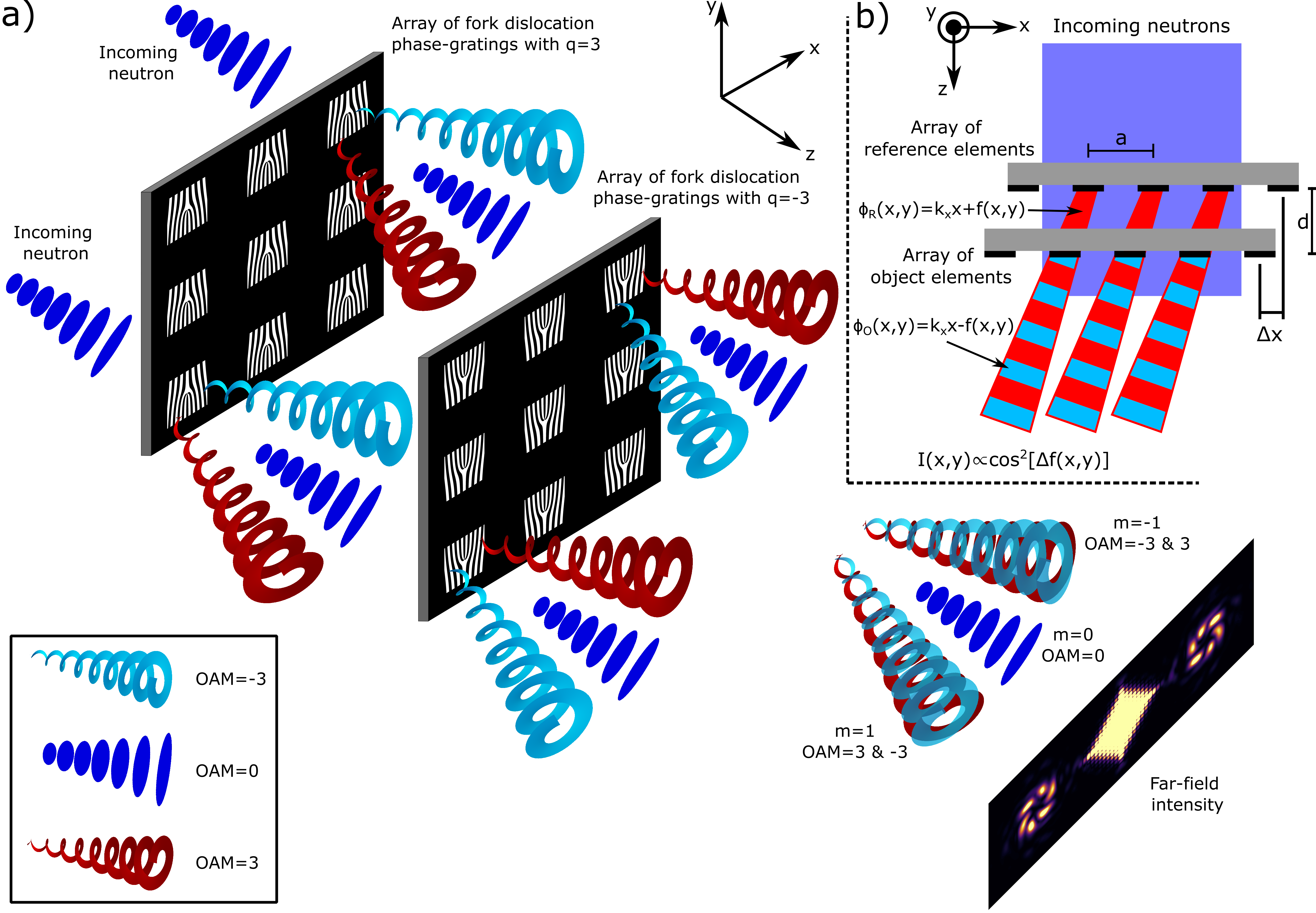}
    \caption{a) Pictorial depiction of the implemented SANS interferometry with  orbital angular momentum (OAM) states. Ref.~\cite{sarenac2022experimental} showed that a single array of fork-dislocation phase-gratings prepares helical waves in the diffraction orders that carry well-defined OAM values and manifest doughnut-like intensity profiles. Here, to observe the transverse phase profile of these helical waves we create a coherent superposition between $\ell$ and $-\ell$ OAM states. Note that we have neglected the higher order interference as the direct beam is orders of magnitude brighter than the diffraction orders. b) The general idea behind the SANS interferometry technique to observe the phase profile of a structured neutron beam with non-trivial propagation characteristics. We consider an incoming neutron with wave vector $k_0$ and two arrays with periodicity $a$ and separation distance $d$. The array of reference beams need to spatially overlap the array of object beams in order to interfere at the camera. Therefore, the amplitude of the reference beams needs to diffract similar to the object beams. One straightforward approach to achieve this is by ensuring that the phase profile of the reference beams $\phi_R(x,y)$ and the object beams $\phi_O(x,y)$ possess a matching carrier wave vector $k_x$, complementary structured phase profiles $f(x,y)$, and enforcing a translational shift $\Delta x$ that ensures overlap (See Eq.~\ref{eq:eq0}).
    }
    \label{fig:concept}
\end{figure*}

\section{SANS interferometry with structured waves}

A conventional Mach-Zehnder (MZ) perfect-crystal neutron interferometer (NI) prepares a coherent superposition of two paths at its output, which provides the phase information through a direct measurement of interference~\cite{Sam}. A distinguishing factor from the standard optical MZ interferometer setups is that the neutron transverse coherence length is orders of magnitude smaller than the beam. Therefore, the action of the MZ NI is applied to each neutron and the observed intensity at the output is their incoherent sum.

\begin{figure*}
    \centering\includegraphics[width=\linewidth]{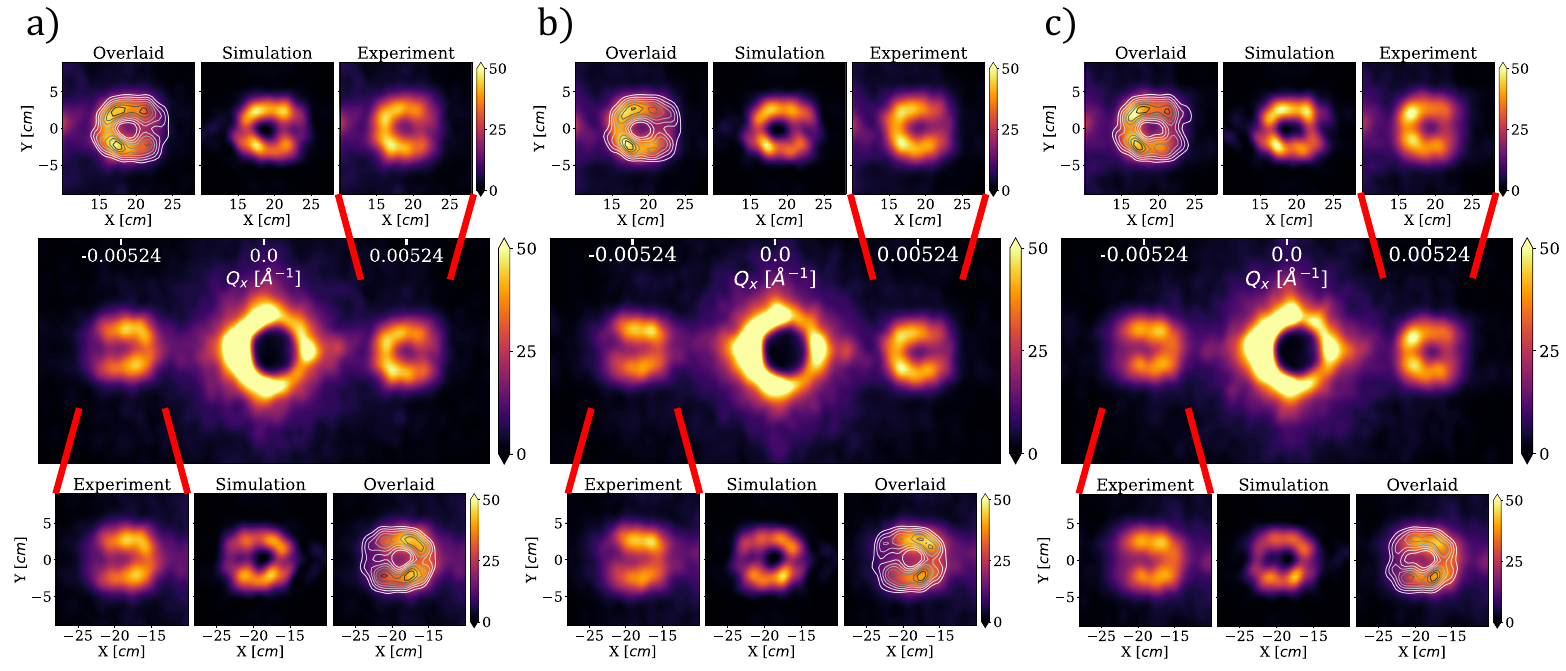}
    \caption{The observed petal-structure signatures of neutron helical wave interference in the first diffraction orders as the pitch and yaw of the double-sided arrays of fork-dislocation phase-gratings was varied. a) We find agreement for a simulation that considers a transverse coherence length of \sigperp$=3~\mu$m and translational offsets between the two arrays of $x_i=0$, $y_i=800$ nm. b) The SANS image after incrementing the yaw by $\approx 0.6$ mrad corresponding to an updated offset of $x_i=0$ nm, $y_i=961$ nm. c) The profile following an additional increment in pitch by $\approx 0.5$ mrad corresponding to $x_i=-135$ nm, $y_i=961$ nm. In this proof of principle experiment, the poor visibility is mainly attributed to the wavelength spread, as depicted on Fig.~\ref{fig:fig3}. 
    }
    \label{fig:data}
\end{figure*}

A challenge arises to integrate interferometry methods and structured neutron waves as their transverse amplitude profiles significantly change with propagation. In order to maximize the interference signal the object beams and the reference beams of the NI need to spatially overlap at the detector plane.  Ref.~\cite{sarenac2022experimental} introduced methods to prepare well-defined OAM states and showed how the amplitude of the neutron wave packets deviates away from their central phase singularity with increasing OAM value. As a result, the radius of the doughnut-like profile of the OAM=3 (OAM=7) states at the neutron camera was measured to be $\approx3$ cm ($\approx5$ cm) which was significantly bigger than the measured half width half maximum, $\approx1$ cm, of the OAM=0 state. 

One way to achieve an appreciable overlap between the reference beams and the object beams is to utilize reference beams that possess complementary spatial phase profiles resulting in similar time-dependent amplitude profiles. In the case of the OAM preparation methods of Ref.~\cite{sarenac2022experimental} this entails creating a coherent superposition between $\ell$ and $-\ell$ OAM states in each diffraction order, as shown in Fig.~\ref{fig:concept}a. The intensity at the camera would take on the form of the well-known petal structure indicative of the OAM~\cite{fickler2013real}, while the OAM dependent phase shift between the two beams would cause the petal profile to rotate. The general concept is shown in Fig.~\ref{fig:concept}b where we consider two complementary arrays of fork-dislocation phase-gratings with periodicity $a$ and separation distance $d$. The interference at the detector is achieved by ensuring that the phase profile of the reference beams $\phi_R(x,y)$ and the object beams $\phi_O(x,y)$ possess matching carrier wave vector $k_x$ and complementary structured phase profiles $f(x,y)$. It follows that there are two interferometric conditions. First, the presence of a translational offset $(\Delta x)$ between the object and reference elements that ensures the reference and object beams are collinear:

\begin{align}
\Delta x\approx d\cdot\tan(k_x/k_0)=d\cdot\tan(\lambda/p)\approx d\lambda/p,
\label{eq:eq0}
\end{align}

\noindent where $\lambda~(k_0)$ is the neutron wavelength (wave vector), and $p~(k_x)$ is the periodicity (wave vector) of the object and reference elements. If $\Delta x$ is greater than the array period $a$, then the effective translation becomes $\Delta x$ mod $a$. Note that the highest contrast occurs for the case where the translational offset is a multiple of the array periodicity, $\Delta x=n\cdot a$ for some integer $n$. The second condition is that the transverse coherence length needs to be greater than $\Delta x$, as the interference is between the first diffraction orders of the incoming beam.

\begin{figure*}
    \centering\includegraphics[width=\linewidth]{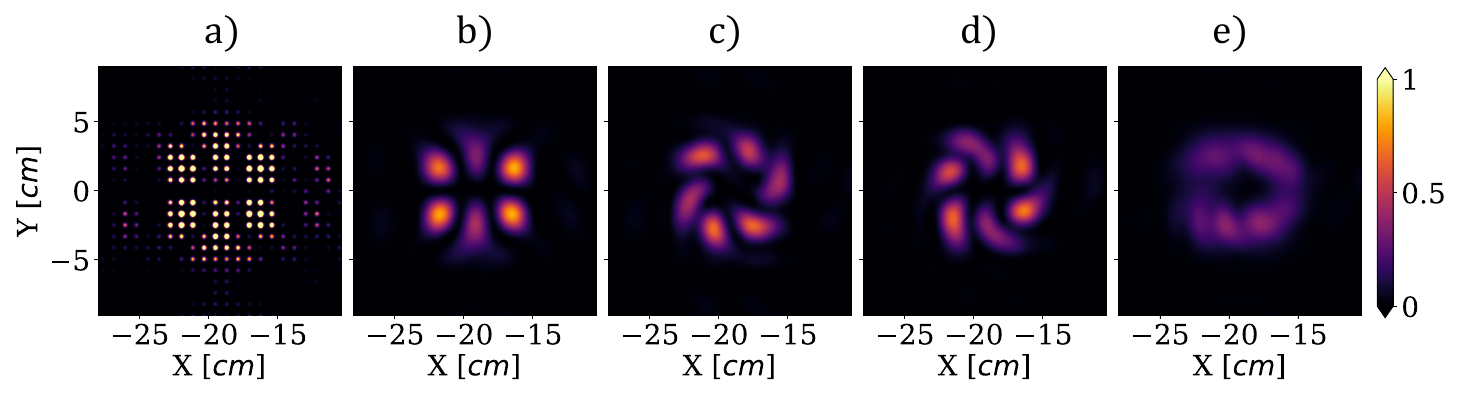}
    \caption{The effect of array structure, distance between the reference and object elements, phase shift between the two interfering OAM states, and wavelength distribution on the petal-structure intensity profile produced in the first diffraction order by the double-sided arrays of fork-dislocation phase gratings. a) Considering an ideal coherent and monochromatic neutron beam incident on object and reference elements with negligible distance between them, one would be able to resolve the interference pattern that is both indicative of the OAM interference and the two-dimensional array structure. b) Removing the higher order frequencies corresponding to the two-dimensional array structure results in the OAM petal-structure being emphasized. c) Increasing the distance between the reference and object elements to the experimental value of $280~\mu$m introduces a phase shift and propagation diffraction that manifests a winding structure in the interference pattern. d) Adding to the object beams an additional OAM dependent $\pi$ phase shift effectively rotates the petal pattern. e) Increasing the wavelength distribution to the experimental $\Delta\lambda/\lambda\approx0.13$ smears the interference pattern.  
    }
    \label{fig:fig3}
\end{figure*}

\section{Materials and Methods}

To aid in alignment we employed double-sided nanofabrication methods. An array of fork-dislocation phase-gratings was fabricated on both sides of a silicon wafer where on one side there was a $q=3$ topology and the other side possessed $q=-3$, which results in them being 180$\degree$ relative to each other. The arrays covered a 0.5 cm by 0.5 cm area and consisted of $6,250,000$ individual 1 $\mu$m by 1 $\mu$m fork-dislocation phase-gratings, where each one possessed a period of $120$ nm, height $500$ nm, and was separated by 1 $\mu$m on each side from the other fork-dislocation phase-gratings. The fab procedure for a single array can be found in the Supplementary Material of Ref.~\cite{sarenac2022experimental}. After fabricating the front side of the wafer, it was spun with a thick PMMA to protect the fabricated gratings for the future steps. Using the back-side alignment feature at a maskless aligner (MLA 150, Heidelberg Instrument~\cite{NIST_disclamer}), cross-shape marks were placed on the other face of the wafer via the e-beam evaporation and lift-off techniques (Cr/Au 20/100 nm). The back-side alignment at MLA has an alignment accuracy of better than 1 $\mu$m. Based on these alignment marks, the second array of gratings was fabricated on the other face of the wafer. 

A preliminary study was done at the SANS-I beamline at the Paul Scherrer Institute~\cite{kohlbrecher2000new}, as a result of which an improved setup was devised and implemented at the GP-SANS beamline at the High Flux Isotope Reactor at Oak Ridge National Laboratory~\cite{wignall201240}. The experimental parameters were the same as those in Ref.~\cite{sarenac2022experimental} which described the observation of neutron helical waves. The double-sided wafer was placed inside a rotation mount which was placed 17.8 m away from a 20 mm diameter source aperture. A 4 mm diameter sample aperture was placed right in front of the sample. The distance from the sample to the camera was 19 m, and the camera size spans an area of $\approx 1$ m$^2$ with each pixel being $\approx 5.5$ mm by $4.3$ mm in size. The wavelength distribution was triangular with $\Delta\lambda/\lambda\approx0.13$ and a central wavelength of $12$~\AA, and the standard deviation of the resolution distribution was estimated to be $\sigma_{Q}=0.00016$~\AA$^{-1}$. The neutron wavelength is selected by a turbine-like velocity selector which eliminates the fractional $\lambda$ contributions that are present when a monochromator crystal is used.

An empty beam scan without a sample and a background scan for a plain silicon wafer of equivalent size and thickness were collected. These measurements were used to take into account the factors that contribute to losses in intensity and increased background scattering noise. The final SANS data was passed though a low-pass filter to remove the Poissonian noise that varies from pixel to pixel.

\section{Implementation and Results}

The silicon wafer sample with the double-sided arrays of fork-dislocation phase-gratings was mounted in a holder that was then placed on a manual tip-tilt stage with access to pitch ($\approx2.9$ mrad/rev) and yaw ($\approx 6.9$ mrad/rev) degrees of rotation. The experiment consisted of measuring the SANS intensity profile for 90 min as the pitch and yaw were systematically varied by $1/6$ rev and $1/12$ rev, respectively. Given the neutron wavelength of $12$~\AA, the phase grating period $p=120$ nm, and the thickness of the wafer $d=280~\mu$m, we can determine the optimal translational offset (see Eq.~\ref{eq:eq0}) to be $\Delta x\approx 2.8~\mu$m; which from the fact that the period of the array is $2~\mu$m reduces to $\Delta x\approx 0.8~\mu$m or a wafer rotation angle of $\approx 3$ mrad.  

To determine the behaviour of the interference pattern at the camera we analyze the propagation of the neutron wave function through the object and reference elements given the experimental parameters. Let us consider the wave function of a neutron wave packet after passing through the first array:

\begin{align}
\psi_i(x,y)=\frac{1}{\sigperpEQ\sqrt{\pi }}e^{-\frac{x^2+y^2}{2\sigperpSqr}}e^{iNb_c\lambda h F_3[x,y]},
\label{eq:eq1}
\end{align}

\noindent where $\sigma_x=\sigma_y=$~\sigperp~is the neutron transverse coherence length, $Nb_c$ is the scattering length density of the sample material, $\lambda$ is the neutron wavelength, $h$ is the height of the grating structures, and $F_q(x,y)$ is the profile of an array of fork-dislocation phase-gratings whose individual profiles are given by:

\begin{align}
	\frac{1}{2}\text{sgn}\left(\text{sin}\left[\frac{2\pi  }{p}x+q\tan^{-1}(y/x)\right]\right)+\frac{1}{2},
	\label{eqn:gratingProfile}
\end{align}

\noindent where $p$ is the grating period and $q$ is its topological charge. Note that for $F_q(x,y)$ in our simulations we considered the SEM profiles found in the Supplementary Material of Ref.~\cite{sarenac2022experimental}. The wavefunction is then propagated a distance of $280~\mu$m as per the wafer thickness, at which point the phase profile corresponding to the second array is added: $Nb_c\lambda h F_{-3}[x-x_i,y-y_i]$, where $\{x_i,y_i\}$ is the translational offset between the two arrays that results from the pitch and yaw increments. The intensity $I(x',y')$ at the camera is well approximated by:

\begin{align}
I\left(x'\equiv k_x\lambda z,y'\equiv k_y\lambda z\right)=|FT\{\psi_f(x,y)\}|^2.
\end{align}

\noindent where $FT\{\}$ is the Fourier Transform, $z$ is the distance from the wafer to the camera, and $\psi_f(x,y)$ is the wave function after the second array. To account for the experimental parameters that set the resolution of $\sigma_{Q}=0.00016$~\AA$^{-1}$, we add a low-pass filter with the corresponding cutoff frequency to the obtained intensity profile. This removes the diffraction signatures from the $2~\mu$m periodicity of the arrays.

Fig.~\ref{fig:data}a shows the observed petal-structure signatures of neutron helical wave interference in the first diffraction orders. Using \sigperp$=3~\mu$m we find good agreement for Fig.~\ref{fig:data}a for $x_i=0$, $y_i=800$ nm. Fig.~\ref{fig:data}b shows the SANS image after incrementing the yaw by $\approx 0.6$ mrad and Fig.~\ref{fig:data}c shows the profile following an additional increment in pitch by $\approx 0.5$ mrad.   The simulated profiles are then determined by the addition of the corresponding translational offsets of $x_i=0$ nm, $y_i=961$ nm for Fig.~\ref{fig:data}b, and $x_i=-135$ nm, $y_i=961$ nm for Fig.~\ref{fig:data}c.

Fig.~\ref{fig:fig3} shows how the petal-structure intensity profile is affected by array periodicity, distance between the reference and object elements, phase shift between the two interfering OAM states, and wavelength distribution. Note that in  Fig.~\ref{fig:fig3} we consider the described experimental parameters but with the mentioned optimal case scenario of array period equaling the transitional offset $a=\Delta x=2.8~\mu$m.

Lastly, we can note that although our model considers the full diffraction spectra, the main contributing factors are the two complementary OAM states as shown in Fig.~\ref{fig:concept}a. Taking note that each wavelength results in a displaced intensity profile at the camera as per the diffraction angle of $\approx \lambda/p$, it can be shown that for the considered parameters the superposition of OAM~$=-3~\&~3$ simulates almost identical petal-structure intensity profiles as shown in Fig.~\ref{fig:data}~\&~\ref{fig:fig3}.

\section{Conclusion and Discussion}

We have introduced and demonstrated a novel neutron interferometry technique for extracting phase information that is typically lost in SANS measurements. By preparing reference beams with complementary spatial phase profiles to that of the object beams we observed the petal-structure signatures of neutron helical wave interference for the first time. With the growing number of neutron-imaging and -scattering user facilities worldwide~\cite{garoby2017progress,alarcon2023fundamental}, the demonstrated techniques are set to make a high impact in the next-generation of neutron scattering studies. Furthermore, it is also possible that this approach could be extended to small--angle x--ray scattering~\cite{PhysRevLett.126.117201}. Lastly, these methods can be extended to implementations with helical waves of atoms and molecules~\cite{luski2021vortex}, which also rely on arrays of fork-dislocation gratings. 

The employed neutron optics element, which was composed of an array of fork-dislocation phase-gratings on both sides of a silicon wafer, adds an additional capability to the toolbox of structured neutron waves for the studies of helical features. Analogous to a $\pi/2$ spin flipper that prepares a coherent superposition of two orthogonal spin states, this element prepares a coherent superposition of two OAM states. Future experiments employing this element will study helical phase shifts from materials. 

There are several ways to improve the observed visibility of the presented helical wave interference. In this proof of principle experiment, the poor visibility is mainly attributed to the width of the wavelength distribution. Furthermore mechanical drifts and ambient conditions, which typically are not a concern for SANS measurements of grating structures, can induce dephasing that degrades the visibility of the interference pattern. We can also note that as nanofabrication methods advance it is expected that a much higher signal in the diffraction orders will become obtainable. For example, if amplitude-gratings or phase-grating with heights of $5~\mu$m were feasible then the intensity in the diffraction orders would be around a hundred times higher compared to this experiment that employed phase-grating heights of 500 nm. The corresponding measuring time of $<1$ min would enable a wide range of possibilities.

\section*{Acknowledgements}

This work was supported by the Canadian Excellence Research Chairs (CERC) program, the Natural Sciences and Engineering Research Council of Canada (NSERC), the Canada  First  Research  Excellence  Fund  (CFREF), and the US Department of Energy, Office of Nuclear Physics, under Interagency Agreement 89243019SSC000025. This work was also supported by the DOE Office of Science, Office of Basic Energy Sciences, in the program ``Quantum Horizons: QIS Research and Innovation for Nuclear Science'' through grant DE-SC0023695. A portion of this research used resources at the High Flux Isotope Reactor, a DOE Office of Science User Facility operated by the Oak Ridge National Laboratory. This work is based partly on experiments performed at the Swiss spallation neutron source SINQ, Paul Scherrer Institute, Villigen, Switzerland.

\bibliography{OAM}

\begin{thebibliography}{47}
\bibitem [{NIS()}]{NIST_disclamer}%
  \BibitemOpen
  \href@noop {} {}\bibinfo {note} {Certain trade names and company products are
  mentioned in the text or identified in an illustration in order to adequately
  specify the experimental procedure and equipment used. In no case does such
  identification imply recommendation or endorsement by the National Institute
  of Standards and Technology, nor does it imply that the products are
  necessarily the best available for the purpose.}\BibitemShut {Stop}%
\end{thebibliography}

\clearpage
\onecolumngrid

\end{document}